\documentclass[journal]{IEEEtran}
\ifCLASSINFOpdf
\else
\fi
\usepackage{graphicx}
\usepackage{amsmath}
\usepackage[utf8]{inputenc}
\usepackage[T1]{fontenc}
\usepackage{mathtools}
\usepackage{booktabs}
\usepackage{multirow}
\usepackage{array}

\newenvironment{conditions}
  {\par\vspace{\abovedisplayskip}\noindent\begin{tabular}{>{$}l<{$} @{${}={}$} l}}
  {\end{tabular}\par\vspace{\belowdisplayskip}}

\begin{document}
%
\title{Cuffless Blood Pressure Estimation from Electrocardiogram and Photoplethysmogram Using Waveform Based ANN-LSTM Network}
%
%
%

\author{\IEEEauthorblockN{
Md. Sayed Tanveer and 
Md. Kamrul Hasan\IEEEauthorrefmark{1}}
\thanks{All authors are with the Department of Electrical and Electronic Engineering, Bangladesh University of Engineering and Technology, Dhaka-1205, Bangladesh.}
\thanks{E-mail: \IEEEauthorrefmark{1}khasan@eee.buet.ac.bd}
}

%
%

\markboth{}%
{Shell \MakeLowercase{\textit{et al.}}: Bare Demo of IEEEtran.cls for Journals}
%



\maketitle

\begin{abstract}
\textit{Goal:} Although photoplethysmogram (PPG) and electrocardiogram (ECG) signals can be used to estimate blood pressure (BP) by extracting various features, the changes in morphological contours of both PPG and ECG signals due to various diseases of circulatory system and interaction of other physiological systems make the extraction of such features very difficult. \textit{Methods:} In this work, we propose a waveform-based hierarchical Artificial Neural Network--Long Short Term Memory (ANN-LSTM) model for BP estimation. The model consists of two hierarchy levels, where the lower hierarchy level uses ANNs to extract necessary morphological features from ECG and PPG waveforms and the upper hierarchy level uses LSTM layers to account for the time domain variation of the features extracted by lower hierarchy level. \textit{Results:} The proposed model is evaluated on 39 subjects using the Association for the Advancement of Medical Instrumentations (AAMI) standard and the British Hypertension Society (BHS) standard. The method satisfies both the standards in the estimation of systolic blood pressure (SBP) and diastolic blood pressure (DBP). For the proposed network, the mean absolute error (MAE) and the root mean square error (RMSE) for SBP estimation are 1.10 and 1.56 mmHg, respectively, and for DBP estimation are 0.58 and 0.85 mmHg, respectively. \textit{Conclusion:} The performance of the proposed hierarchical ANN-LSTM model is found to be better than the other feature engineering-based networks. It is shown that the proposed model is able to automatically extract the necessary features and their time domain variations to estimate BP reliably in a noninvasive continuous manner. \textit{Significance:} The method is expected to greatly facilitate the presently available mobile health-care gadgets in continuous BP estimation.
\end{abstract}


%
\IEEEpeerreviewmaketitle

\section{Introduction}
Blood pressure (BP) is the pressure exerted by the circulating blood on the blood vessels. It is an important physiological parameter which is widely used by physicians to check the condition of the circulatory system of patients. Blood pressure is important because the higher the blood pressure is, the higher is the risk of health problems in the future. High blood pressure, or hypertension, puts extra strain on the blood vessels and causes damages to internal organs such as heart, brain, eyes, kidneys etc. Chronic hypertension can lead to heart disease, stroke, kidney failure, blindness, rupture of blood vessels, premature mortality and disability \cite{world2013global}. Such a dangerous medical condition is actually very common in both developing countries and industrialized nations. Based on the World Health Organization report, 24\% of men and 20.5\% of women are subject to hypertension at various degrees \cite{world2015world}. Low blood pressure, or hypotension, can also be dangerous to the patients. Severely low blood pressure can cause oxygen and nutrient deprivation in the brain and other vital organs, resulting in a serious medical condition called shock \cite{silverman2005shock}.

The normal BP ranges for systolic blood pressure (SBP) and diastolic blood pressure (DBP) in adults are considered to be 90-129 mmHg and 60-84 mmHg, respectively \cite{gabb2016guideline}. SBP and DBP lower than this range indicates hypotension. Likewise, SBP and DBP higher than this range indicates hypertension. SBP and DBP in subjects do not always remain unchanged. Various factors such as food, physical activity, psychological state may influence both SBP and DBP, which may lead a subject to BP measurements outside of the normal range. This variability in BP necessitates the continuous measurement of BP. In recent studies, high blood pressure variability has been shown to be a significant predictor for stroke \cite{rothwell2010prognostic}. Thus, for accurate diagnosis and treatment of both hypertension and hypotension, and for early detection of life threatening situations such as stroke or shock, continuous BP measurement is essential.

The conventional automated blood pressure measurement devices use the cuff-based oscillometric \cite{geddes1982characterization} and auscultatory \cite{geddes1966introduction} methods. The repeated measurement operation of such devices is discontinuous in nature, with intervals between measurements greater than at least two minutes, and is reported to be uncomfortable by many users. And for these reasons, cuff-less implementation of such devices has become more and more desirable. Pulse wave velocity (PWV) \cite{landowne1957method} has been proved to be a promising parameter for continuous BP measurement. Pulse transit time (PTT) \cite{geddes1981pulse}, the time needed for the pulse wave generated at heart to propagate to a peripheral site of the body, is found to be inversely related to PWV and thus, it has gained the interest of many researchers. Many works have tried to fit various regression-based models to estimate BP using PTT, but failed to obtain satisfactory results \cite{poon2006cuff}, \cite{kumar2014cuffless}.

Other than PTT, many other morphological features, that can be extracted from PPG signals, have also been found to be correlated with blood pressure \cite{elgendi2012analysis}. These features show good correlation with the blood pressure of an individual. Combined with the features extracted from PPG signals, PTT is able to provide more accurate estimation of the blood pressure. In recent years, Kachuee \textit{et al.} \cite{kachuee2017cuffless} and Xu \textit{et al.} \cite{xu2017continuous} have used PTT as the main feature along with many other parameters extracted from PPG signals as auxiliary features and have used machine learning algorithms to obtain models with better accuracy and more promising results than PTT-only models \cite{poon2006cuff}, \cite{kumar2014cuffless}.

As the ECG and PPG signals can get affected by various diseases, drugs and other outside influences \cite{takazawa1998assessment}, \cite{opie2004heart}, \cite{selvaraj2008assessment}, extracting the features can often be quite difficult. Automatic extraction of necessary features from PPG signals is thus a necessity. Kachuee \textit{et al.} \cite{kachuee2017cuffless} used a time domain approach to model the PPG waveforms using traditional machine learning algorithms such as decision tree and boosting. Xing \textit{et al.} \cite{xing2016optical} used a frequency domain approach to extract the necessary features from PPG waveforms. These methods do not provide any significant improvement in performance compared to the feature-engineering based methods. Recent studies have shown that neural networks can extract the necessary features automatically without any complex feature engineering \cite{mao1995artificial}.

The above discussed models do not account for the variations in the extracted features with respect to time. As the arterial pressure of the human body is regulated by the autonomic nervous system and renal-body fluid mechanism that involve multiple feedback control loops \cite{guyton1972arterial}, \cite{mancia2014autonomic}, its effects can also be seen in the ECG and PPG signals as modulations \cite{charlton2016assessment}. The estimation accuracy of BP can be increased by accounting for not only the relevant features, but also their time domain variations. For modeling complex time series, Long Short Term Memory (LSTM) \cite{hochreiter1997long} has been found to be extremely useful. Su \textit{et al.} \cite{su2017predicting} has shown that blood pressure estimation accuracy can be improved by using deep LSTM networks by using hand-crafted features.

In this work, we propose a waveform-based ANN-LSTM model that extracts necessary features automatically using ECG and PPG waveforms. The model consists of a hierarchical structure where the lower hierarchy level uses ANNs to extract the necessary features from the ECG and PPG waveforms and the upper hierarchy level uses stacked LSTM layers to learn the time domain variations of the features extracted in the lower hierarchy level. The proposed model is able to estimate BP with high accuracy in an end-to-end manner.

The paper is organized as follows: Section II discusses the theories behind the BP estimation method. Section III discusses various limitations and problems of extracting features from PPG signals. Section IV discusses the proposed method in more details. Section V illustrates the experimental results and compares with other algorithms in this field. Lastly, section VI concludes the paper.

\section{Background}
\begin{figure}[t]
\centering
\includegraphics[width=\linewidth]{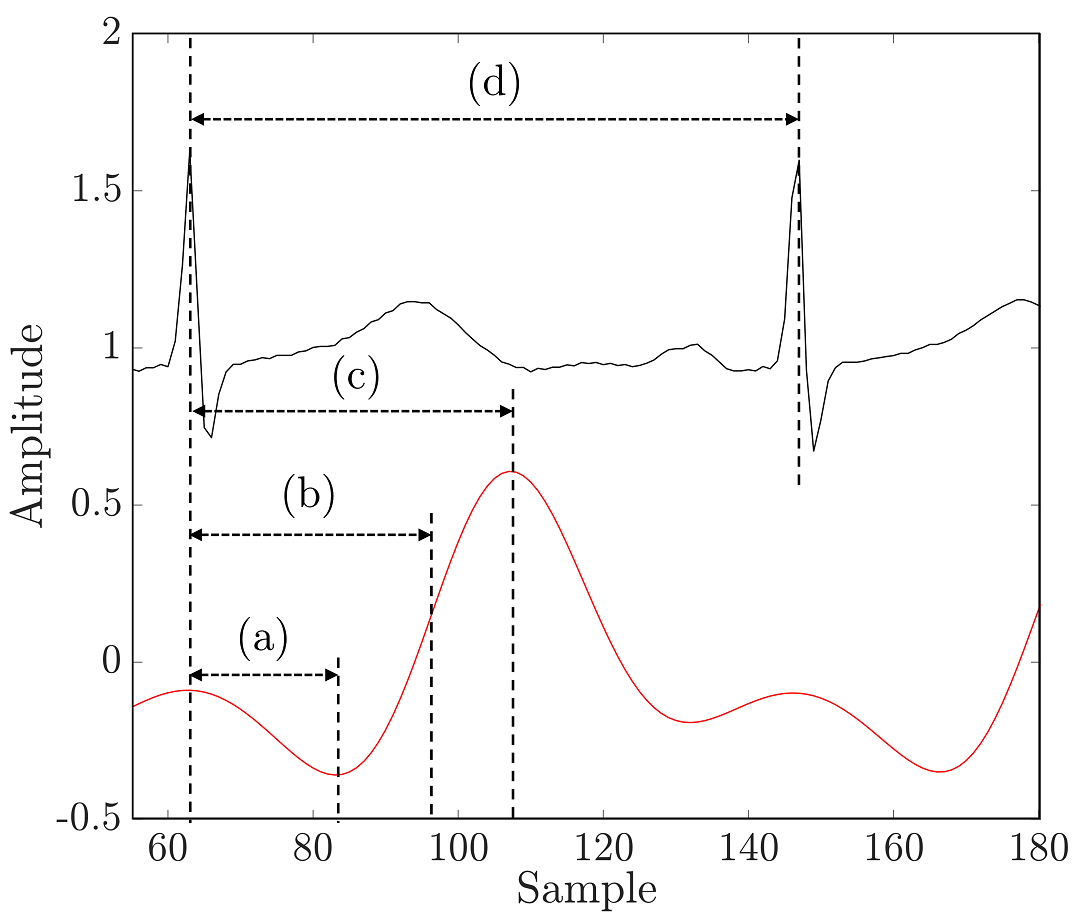}
\centering
\caption{Measurement PTT from ECG R peak to the (a) onset of the PPG waveform, (b) peak of 1st derivative of the PPG waveform and (c) peak of PPG waveform. Here, (d) indicates the R-R interval, which is the time interval between two consecutive heart beats.}
\label{fig01_ptt}
\end{figure}

ECG signal is the electric potential difference generated across the heart of a subject obtained using electrodes. On the other hand, PPG signal is the volumetric measurement of blood at the peripheral site of a body such as earlobe, fingertip etc obtained from measuring the optical signal transmitted through or reflected from the subject's tissue \cite{kamal1989skin}. After ventricular contraction of heart, which is indicated by a sharp peak in the ECG signal, blood is pumped to every part of the body. The pressure generated due to the ventricular contraction of heart propagates through the blood vessels to the peripheral sites as the blood pumped out of the heart increases the volume of the blood in the blood vessels. The extra blood generates pressure on the vessel wall and the vessels expand in diameter. This change in blood volume is reflected in the PPG waveform. The time difference between the generation of pressure at heart and the increase in blood volume at the peripheral site is denoted by pulse transit time (PTT), as seen in Fig. \ref{fig01_ptt}, and the velocity of the propagation of this pressure is pulse wave velocity (PWV). The relation between PTT and PWV can be expressed as
\begin{equation}
PWV=\frac{d}{PTT}
\end{equation}
\\
where $d$ is the arterial distance traveled by the pressure wave.
PWV is related to the elastic modulus of the vessel wall by the Moens-Kortweg equation \cite{vlachopoulos2011mcdonald}:
\begin{equation}
c=\sqrt{\frac{Eh}{\rho d}}
\end{equation}
where
\begin{conditions}
	c & pulse wave velocity (PWV) \\
	E & elastic modulus of vessel wall \\
	h & vessel thickness \\
	\rho & blood density and \\
	d & arterial diameter.
\end{conditions}
The elastic modulus of the vessel wall, in turn, is related to the arterial pressure by the equation
\begin{equation}
E=E_0e^{\gamma P}
\end{equation} 
where $E_0$ and $\gamma$ are subject-specific parameters \cite{hughes1979measurements}. The equations (1)$-$(3) establish a relation between BP and PTT. Besides PTT, many other parameters extracted from PPG waveforms have been found to be correlated with BP \cite{elgendi2012analysis}. These parameters have been found to be indicators of the systemic vascular resistance and arterial compliance, both of which can be modeled to have nonlinear relations with blood pressure according to two-element Windkessel model \cite{westerhof2009arterial}. The PPG-based parameters can be listed as follows:
\begin{figure}[t]
\centering
\includegraphics[width=\linewidth]{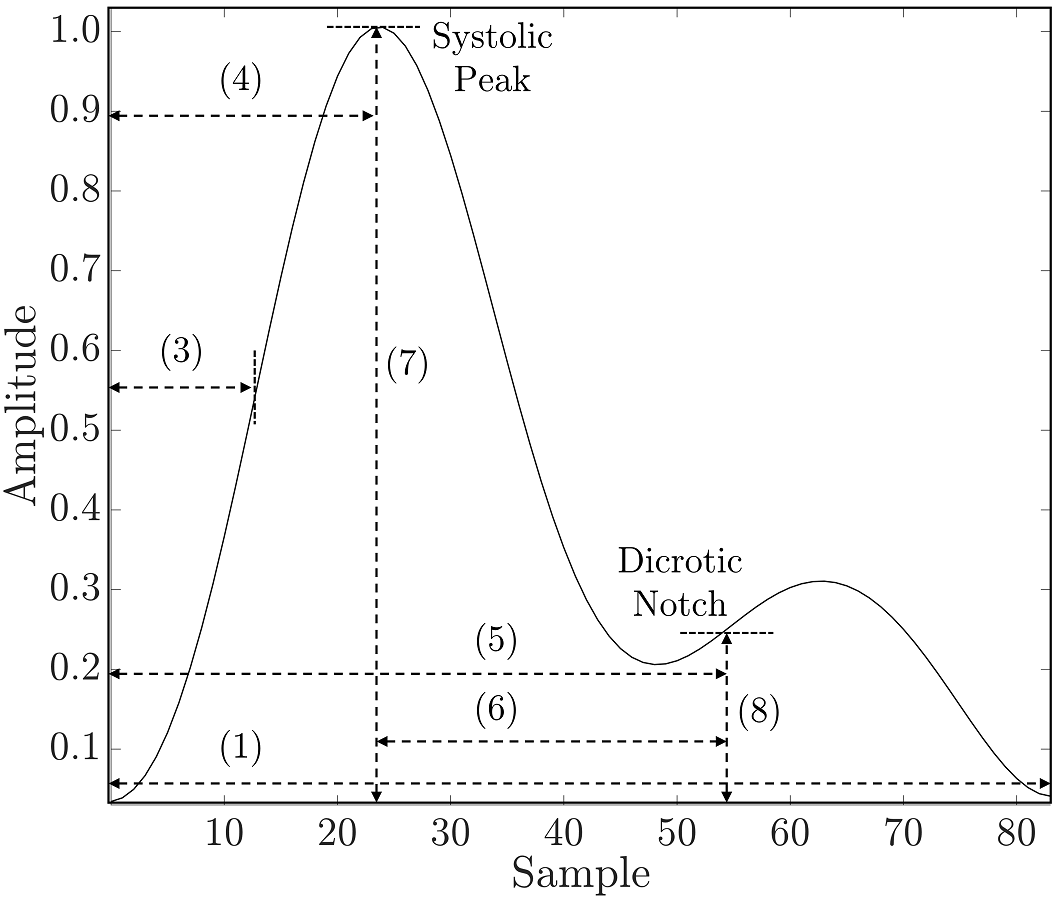}
\centering
\caption{PPG-based morphological features for BP estimation. The numbers indicate the position of the features in the list.}
\label{fig02_ppgfeatures}
\end{figure}
\begin{enumerate}
\item Pulse Interval, the distance between starting and ending points
\item Distance between the adjacent peaks, inversely related to heart rate (HR) \cite{jubadi2009heartbeat}
\item Distance between the starting point and the peak of the derivative
\item Distance between the starting point and the systolic peak, also known as crest time \cite{alty2007predicting}
\item Distance between starting point and the dicrotic notch
\item The distance between systolic peak and dicrotic notch, inversely related to Large Arterial Stiffness Index (LASI) \cite{millasseau2002determination}
\item Amplitude of the systolic peak \cite{chua2010towards}
\item Amplitude of the dicrotic notch
\item Reflection Index (RI), the ratio of systolic peak amplitude to dicrotic notch amplitude \cite{padilla2006assessment}
\item Area under curve (from starting point to the dicrotic notch)
\item Area under curve (from the dicrotic notch to the end point)
\item Inflection point area (IPA) ratio, the ratio of the above mentioned areas \cite{wang2009noninvasive}
\item Ratios of various peaks and notches of the 2nd derivative of PPG waveform \cite{takazawa1998assessment}
\end{enumerate}
and many other features can be calculated from the PPG waveform. Features can also be obtained from various linear and non-linear combinations of the above listed features along with PTT.

\section{Problem Insights}
The proposed model encourages automatic extraction of features from ECG and PPG waveforms because manual feature extraction from ECG and PPG signals is complex and often erroneous. Both ECG and PPG waveforms can get affected by various diseases, drugs and other outside influences \cite{takazawa1998assessment}, \cite{opie2004heart}, \cite{selvaraj2008assessment}. The problems that arise in the manual feature extraction from ECG and PPG signals are discussed in the following.

\subsection{Feature extraction from ECG}
\begin{figure}[t]
\centering
\includegraphics[width=\linewidth]{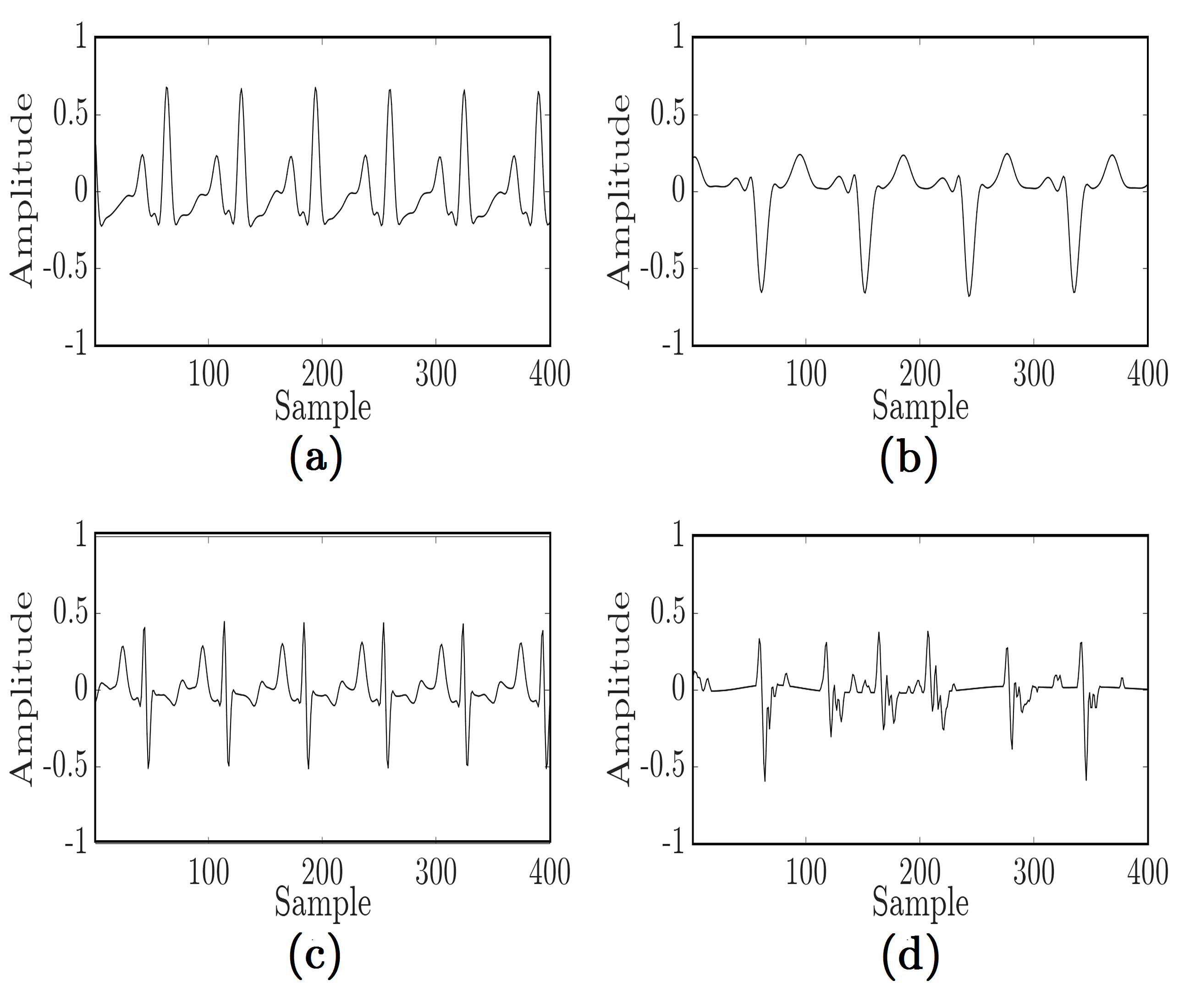}
\centering
\caption{Typical variations in lead II ECG signals. Here, (a) has a sharp positive R peak, whereas (b) has a negative R peak. Both (c) and (d) has sharp deflections in both directions, but the ECG signal in (d) has high irregularity.}
\label{fig03_ecgcontours}
\end{figure}

The primary reason of using ECG is to obtain the PTT measurement. For proper measurement of PTT, proper detection of R peaks is necessary. The contour of the ECG waveform is dependent on the coronary blood flow as well as the placement of the electrodes \cite{opie2004heart}.
Fig. \ref{fig03_ecgcontours} shows lead II ECG recordings of different patients. The patient of the recording in Fig. \ref{fig03_ecgcontours}(a) was not diagnosed with any serious heart condition. The patient of the recording in Fig. \ref{fig03_ecgcontours}(b), on the other hand, was diagnosed with angina. The patient with the recording of \ref{fig03_ecgcontours}(c) had coronary bypass surgery. The patient with the abnormal ECG of Fig. \ref{fig03_ecgcontours}(d) was admitted in the ICU due to heavy bleeding. The contours in Fig. \ref{fig03_ecgcontours} are not the only obtainable waveform shapes, there are other varieties with different level of difficulties in them. Any ECG R peak detection algorithm should be able to deal with all sorts of contingencies.

\subsection{Feature extraction from PPG}
\begin{figure}[t]
\centering
\includegraphics[width=\linewidth]{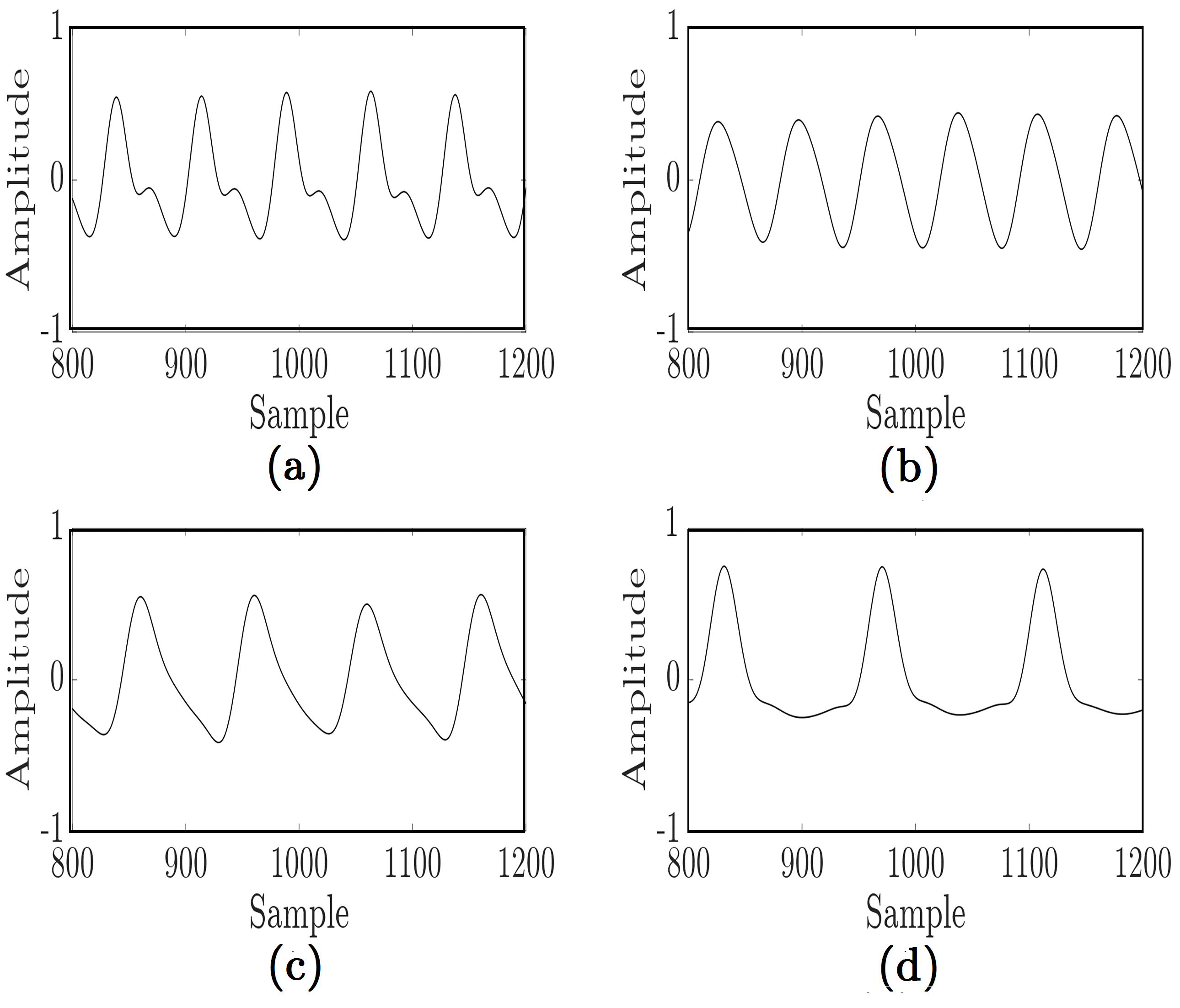}
\centering
\caption{Typical variations in PPG waveforms. Here, (a) has distinct dicrotic notches and thus the relevant features can be extracted fairly easily, (b) shows slight changes in the diastolic (decaying) section of the waveform and the position of the dicrotic inflection point can still be estimated, (c), on the other hand, does not have any visible notch so the relevant features can not be extracted, (d) also does not have any visible notch and the diastolic portions of the waveforms decay much faster than the other three waveforms.}
\label{fig04_ppgcontours}
\end{figure}

An algorithm-based extraction of the relevant features can be very difficult as the waveform contour of the PPG varies from subject to subject. The detectability and accuracy of the above features depends on the PPG waveform contour. In Fig. \ref{fig04_ppgcontours}, we show some PPG waveforms with different contour variations. The contour in Fig. \ref{fig04_ppgcontours}(a) is the ideal PPG contour, where the dicrotic notch can be observed distinctively. All the features listed above can be extracted from this kind of contour. Subjects that are young in age and free from any cardiovascular diseases tend to have this kind of contour. In Fig. \ref{fig04_ppgcontours}(b), the dicrotic notch is not quite distinct, but it can be estimated from the inflection point of the diastolic (downward) portion of the waveform. The detection accuracy is dependent on the PPG waveform constructed after preprocessing. The extraction of feature no. 5-6 and 8-13 from the PPG feature list of section II can be erroneous. For the contour shape of Fig. \ref{fig04_ppgcontours}(c), the same features cannot be extracted since the dicrotic notch is almost nonexistent. The contour shape of Fig. \ref{fig04_ppgcontours}(d) has no visible dicrotic notch either and the diastolic (downward) portion decays much faster compared to other contours, making the feature extraction even harder. The contours in Fig. \ref{fig04_ppgcontours} are not the only obtainable waveform shapes, there are other varieties with different level of difficulties in them. It is very hard to design algorithms that can obtain features from all kinds of PPG waveforms.

\section{Materials and Methods}
In this section, we illustrate the processes which are used for data collection, preprocessing and end-to-end machine learning-based BP estimation.
\\The process flow is as follows:
\begin{itemize}
\item Data Source
\item Preprocessing
\item Proposed Method
\end{itemize}
\subsection{Data Source}
In this paper, Physionet's Multiparameter Intelligent Monitoring in Intensive Care (MIMIC) I database has been used to collect the ECG, PPG and corresponding arterial blood pressure (ABP) signal \cite{goldberger2000physiobank}. Data from 39 patients in ICU belonging to different age groups and sex have been collected and preprocessed. The ECG, PPG and ABP signals were sampled at a sampling frequency of 125Hz. Although MIMIC I database contains data records from 93 patients, only the patients having simultaneous ECG, PPG and ABP recordings are selected for this work. Table I shows a brief description of the patients. From Table I, it is evident that the data is biased towards patients within the age group of 60-80 years. The standard deviations of SBP within the age group of 40-60 years and 60-80 years are high relative to the respective standard deviations of DBP.

The patients were diagnosed with various clinical condition such as respiratory failure, congestive heart failure, sepsis, myocardial infarction, angina, brain injury, cord compression, internal hemorrhage, for example. The patients are subjected to various vasoactive drugs which are known to change the PPG waveform contour. There are also patients with cardiac disorders and thus ECG may deviate from normal sinus rhythm (NSR). In case of ECG lead selection, lead II signal is given priority and for the patients with no lead II recording, lead III or V is selected.
\begin{table}[t]
\begin{tabular}{@{}cccccc@{}}
\multicolumn{6}{c}{\textbf{\begin{tabular}[c]{@{}c@{}}Table I\\ Patient Data Summary Collected from MIMIC I Database\end{tabular}}} \\ \midrule\midrule
\begin{tabular}[c]{@{}c@{}}Age\\ Group\\ (years)\end{tabular} & \begin{tabular}[c]{@{}c@{}}Number\\ of\\ Patients\end{tabular} & \begin{tabular}[c]{@{}c@{}}Number\\ of\\ Male\\ Patients\end{tabular} & \begin{tabular}[c]{@{}c@{}}Number\\ of\\ Female\\ Patients\end{tabular} & \begin{tabular}[c]{@{}c@{}}SBP\\ Mean$\pm$SD\\ (mmHg)\end{tabular} & \begin{tabular}[c]{@{}c@{}}DBP\\ mean$\pm$SD\\ (mmHg)\end{tabular} \\ \midrule
20-40 & 02 & 02 & 00 & 119.77$\pm$12.28 & 77.40$\pm$6.82 \\
40-60 & 03 & 02 & 01 & 137.52$\pm$26.14 & 67.85$\pm$9.13 \\
60-80 & 27 & 14 & 13 & 120.82$\pm$30.35 & 59.12$\pm$11.89 \\
80-100 & 07 & 03 & 04 & 111.29$\pm$15.28 & 57.70$\pm$8.27 \\ \bottomrule
\end{tabular}
\end{table}
\subsection{Preprocessing}
To remove baseline wandering and noise, the windowed signals are bandpass filtered using Tunable-$Q$ wavelet transform (TQWT) \cite{selesnick2011wavelet}. The primary advantage of the TQWT is that it provides all the advantages of wavelet denoising \cite{unser1996review} along with its tunability of the basis wavelet. In TQWT, discrete wavelet transform (DWT) is performed using the Daubechies basis wavelet generated using the provided $Q$ value. By changing the $Q$ value, the wavelet decomposition subbands can be shifted in the frequency domain, which can be used to remove the baseline wandering effectively while preserving the morphological contour shape as much as possible. The $Q$ selection and filtering process is explained in Fig. \ref{fig05_preprocessflow}.
\begin{itemize}
\begin{figure}[t]
\centering
\includegraphics[width=\linewidth]{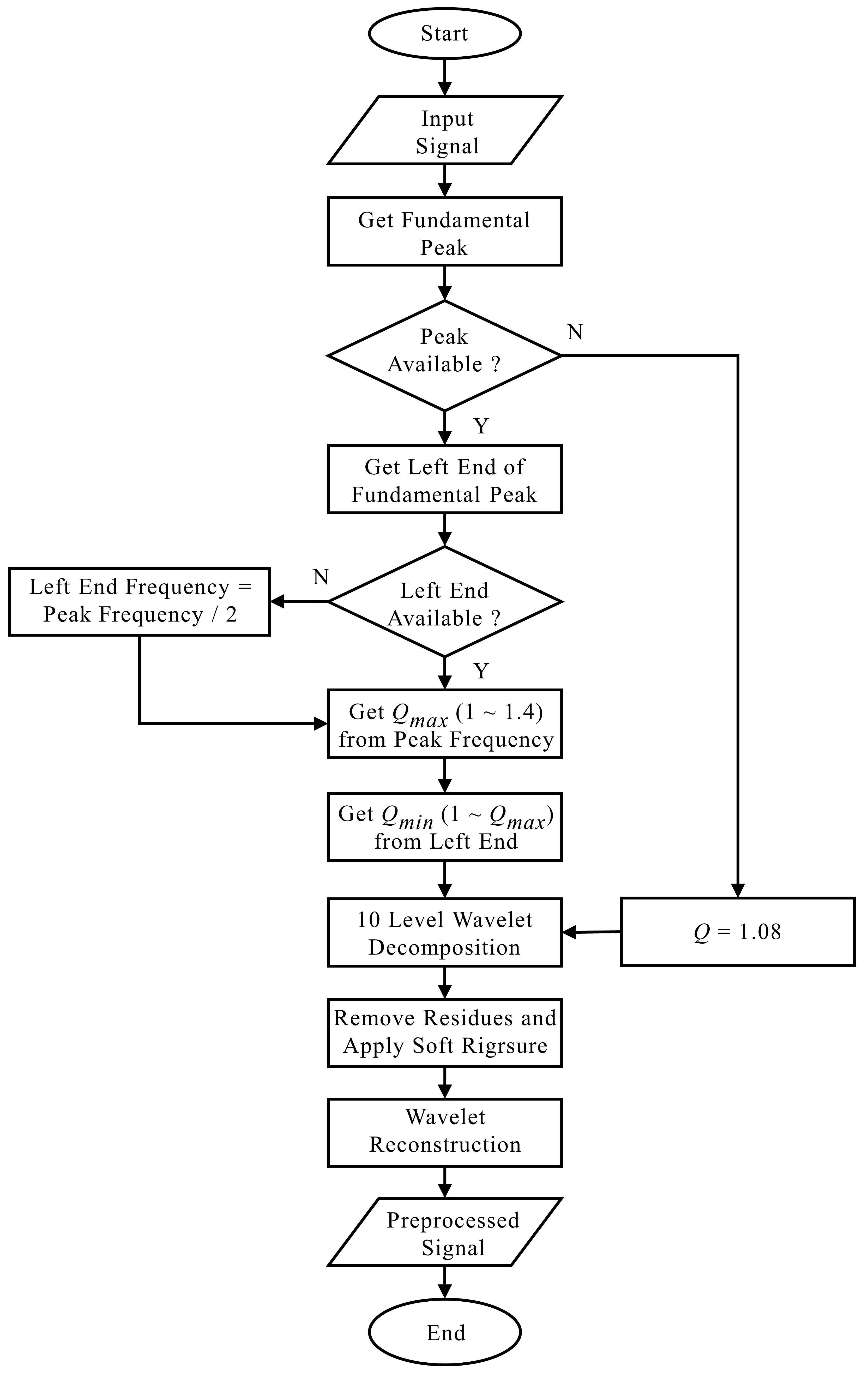}
\centering
\caption{Flow chart of the preprocessing method. Same method is used for both ECG and PPG signals.}
\label{fig05_preprocessflow}
\end{figure}
\item Firstly, Fourier transform is performed on the windowed ECG or PPG signal and then the amplitude is normalized.
\item Then the first peak greater with a prominence greater than 0.4 within the frequency range 1.0-3.5 Hz (if exists) is selected as the fundamental peak.
\item The local minima (if exists) right before the peak is selected as the left end of the peak. If no local minimum is found, then the left end is simply selected as half of the fundamental frequency.
\item $Q$ value is selected from a lookup table using the fundamental peak frequency and the frequency of the left end. The lookup table consists of the center frequency and lower 3 dB cutoff frequency of the 10th level subband for $Q$ values ranging within 1.0-1.4. For $Q$ = 1.0, the center frequency and lower 3 dB cutoff frequency for the 10th level subband are 0.8129 Hz and 0.4309 Hz, respectively and for $Q$ = 1.4, the center frequency and lower 3 dB cutoff frequency for the 10th level subband are 1.9491 Hz and 1.3397 Hz, respectively.
\item The $Q$ value that provides a center frequency closest to the fundamental peak is selected as the upper limit $Q_{max}$ of $Q$ values. Then another $Q$ value is selected within the range 1.0-$Q_{max}$ that corresponds to the 3 dB cutoff frequency closest to the left end frequency. This $Q$ value is our desired $Q$ value. In this way, the fundamental peak is not attenuated and the amplitude of the ECG and PPG signal is preserved.
\item If no fundamental peak is detected, a $Q$ value of 1.08 is selected. For $Q$ = 1.08, the center frequency and lower 3 dB cutoff frequency for the 10th level subband are 1.0020 Hz and 0.5735 Hz, respectively. In such cases, it is highly probable that the ECG or PPG signal is highly corrupted by artifacts and most of the features that require existance of proper waveforms may not be extractable.
\item The residual signal that contains the dc component and baseline wandering is removed.
\item Wavelet denoising using soft Rigrsure thresholding \cite{donoho1995noising} is performed.
\end{itemize}
Finally, we obtain ECG and PPG signals that are free of both baseline wandering and high frequency noise.

\subsection{Proposed Method}
\begin{figure*}[t]
\centering
\includegraphics[width=0.9\textwidth]{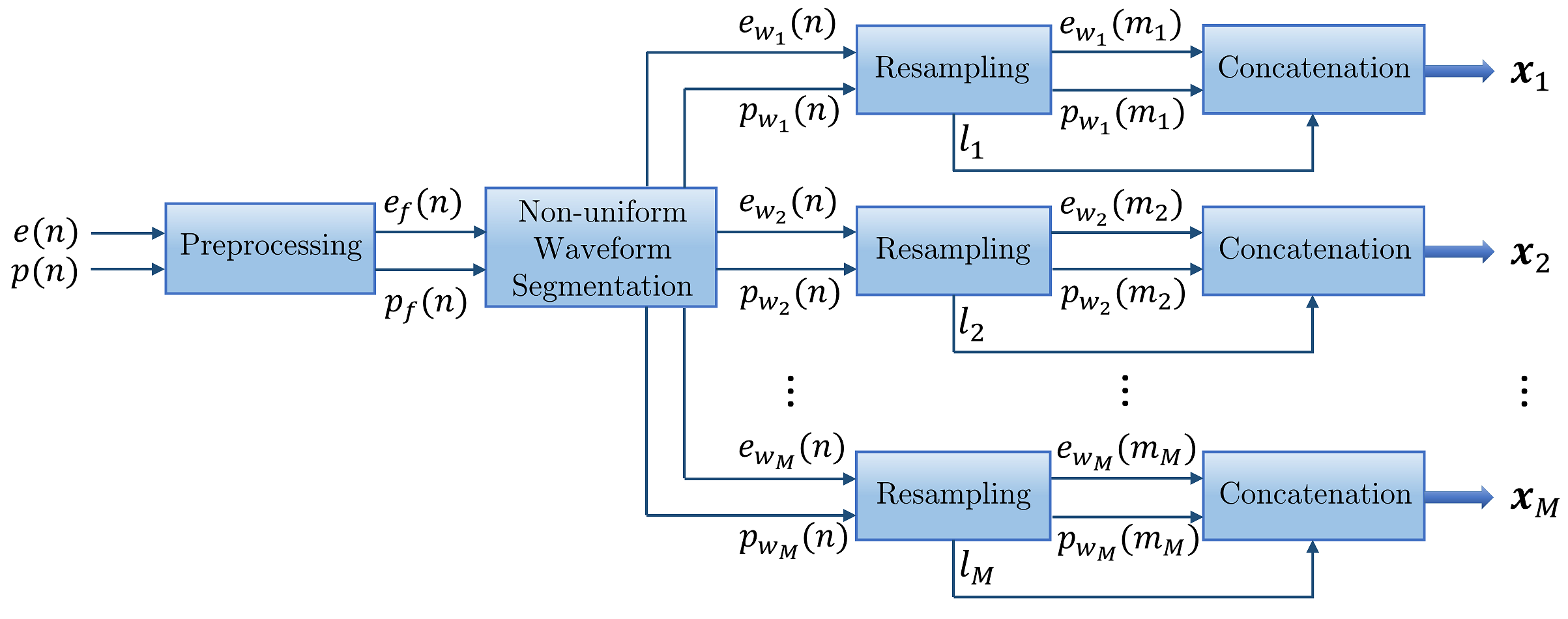}
\centering
\caption{Block diagram of the proposed model.}
\label{fig07_systemsummary}
\end{figure*}

\begin{figure*}[t]
\centering
\includegraphics[width=0.7\textwidth]{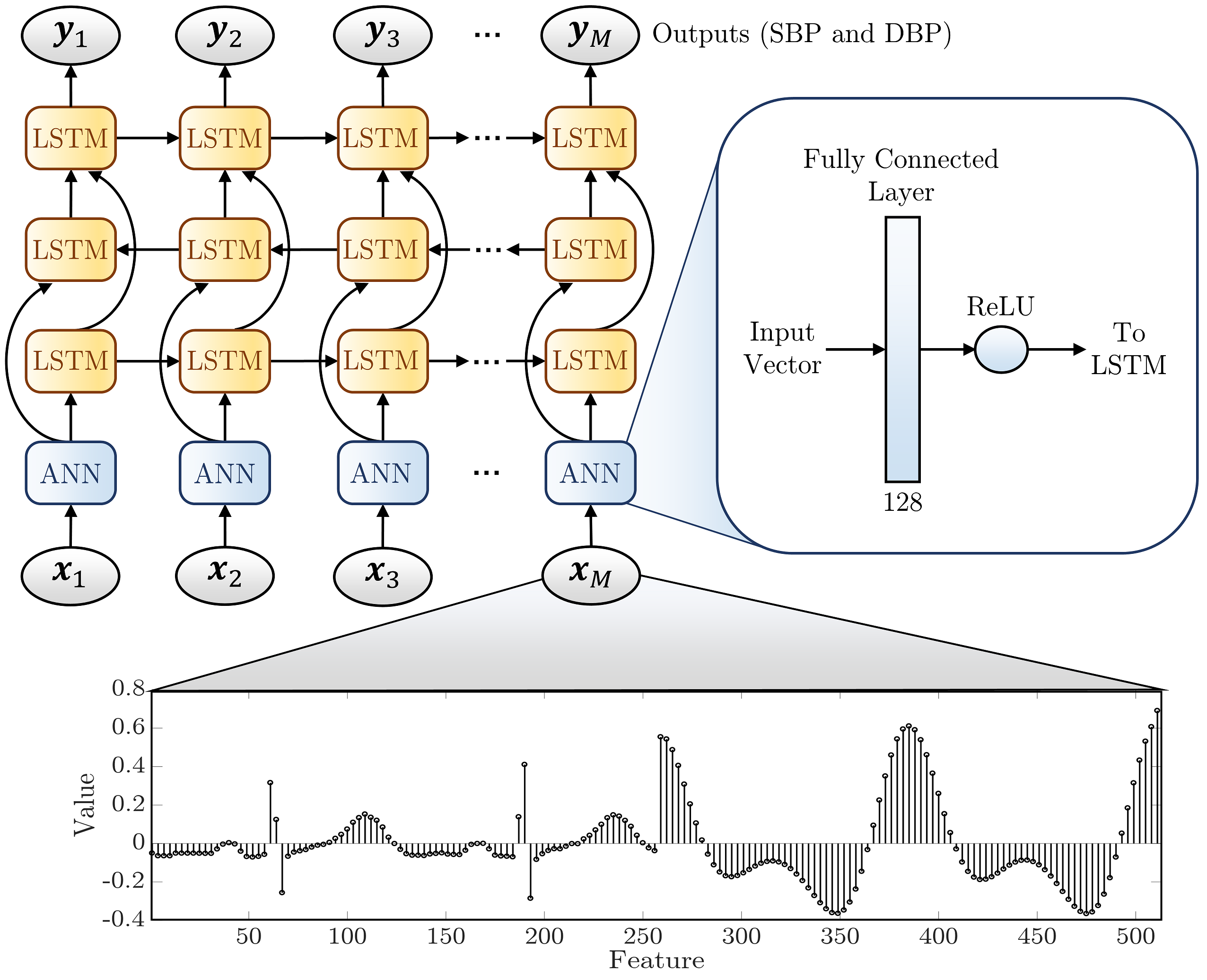}
\centering
\caption{Outline of the proposed hierarchical LSTM network. Here, the fully connected ANN networks extract the necessary features from the feature vector and the LSTM layers learn the variation of the said features.}
\label{fig08_neural}
\end{figure*}

The proposed ANN-LSTM model consists of two hierarchy levels. The lower hierarchy level uses ANNs to learn the features from single instances of ECG and PPG waveforms concatenated together and the upper hierarchy LSTM level learns the temporal relations amongst the features extracted in the lower hierarchy level.  Fig. \ref{fig07_systemsummary} shows a pipeline of the input data preparation process, including preprocessing, and Fig. \ref{fig08_neural} shows an outline of the proposed neural network model.

\subsubsection{Non-uniform Waveform Segmentation}
The proposed algorithm encourages the segmentation of PPG and ECG waveforms for training the neural network. The primary reason behind this segmentation is the varying frequency of the ECG and PPG signals. Heart rate can be less than 60 bpm (for patients of Bradycardia) and can rise up to 200 bpm due to strong sympathetic nervous stimulation \cite{hall2010guyton}. If a constant window of some length, for example, 2s (250 samples) is taken, the number of cycles of the waveforms within that window may vary within 2-6 cycles. This varying number of cycles will create bias in the extracted features towards subjects with higher HR. For this reason, a variable window with the length of three consecutive ECG or PPG peaks (two cycles) is applied and then resampled to 256 samples. Zero padding is avoided since, in the worst case scenario, at least two-thirds of the padded signal will contain zeros which will affect the network negatively.

For training the network, the input waveforms are prepared and then fed to the neural network. The input data preparation process is described in the following.
\begin{itemize}
\item Firstly, PPG segment with a length of three consecutive systolic peaks (two cycles) $p_{w_i}$($n$), $i$ = 1, 2, $\cdot\cdot\cdot$ $M$, where $M$ is the number of the sequential waveform segments, and the corresponding ECG segment $e_{w_i}$($n$), $i$ = 1, 2, $\cdot\cdot\cdot$ $M$, is extracted from the preprocessed PPG and ECG signals $p_f$($n$) and $e_f$($n$), respectively. It is also acceptable to take ECG segment of a length of three R peaks and the corresponding PPG signal instead. In this work, the first approach has been followed.
\item Secondly, the segment length $l_i$, $i$ = 1, 2, $\cdot\cdot\cdot$ $M$, is calculated from the extracted ECG and PPG waveform and is divided by 256 to normalize.
\item Finally, the ECG and PPG signals are resampled to lengths of 256 samples. Concatenating the ECG and PPG resampled waveforms $e_{w_i}$($m_i$) and $p_{w_i}$($m_i$), and the normalized segment length $l_i$ together, a vector $\textbf{x}_i$ of 513 features is obtained. This vector is the waveform-based feature vector.
\end{itemize}
The proposed model requires a sequence of $M$ consecutive feature vectors ${\begin{bmatrix}\textbf{x}_1^T~\textbf{x}_2^T~\cdot\cdot\cdot~\textbf{x}_M^T\end{bmatrix}}^T$. The sequence is prepared as follows:
\begin{itemize}
\item The above process is executed for $M$ sets of three consecutive peaks offset by one peak, and $M$ feature vectors are formed. The peaks are selected as follows: 0-1-2, 1-2-3, 2-3-4 up to ($M-$1)-$M$-($M+$1).
\item Such $M$ feature vectors are extracted to create a sequence of feature vectors for the ANN-LSTM network. Then the corresponding sequence vector ${\begin{bmatrix}\textbf{y}_1^T~\textbf{y}_2^T~\cdot\cdot\cdot~\textbf{y}_M^T\end{bmatrix}}^T$ consisting of SBP and DBP are taken as the output.
\end{itemize}
\begin{table*}[t]
\centering
\begin{tabular}{ccccccccccccc}
\multicolumn{13}{c}{\textbf{\begin{tabular}[c]{@{}c@{}}Table II\\ Comparative Analysis of Different Models for Blood Pressure Estimation\end{tabular}}} \\ \hline\hline
\multicolumn{1}{c|}{\multirow{3}{*}{\textbf{Model}}} & \multicolumn{4}{c|}{\textbf{\begin{tabular}[c]{@{}c@{}}Sequence\\ Length\\ M=1\end{tabular}}} & \multicolumn{4}{c|}{\textbf{\begin{tabular}[c]{@{}c@{}}Sequence\\ Length\\ M=10\end{tabular}}} & \multicolumn{4}{c}{\textbf{\begin{tabular}[c]{@{}c@{}}Sequence\\ Length\\ M=32\end{tabular}}} \\ \cline{2-13} 
\multicolumn{1}{c|}{} & \multicolumn{2}{c|}{\textbf{\begin{tabular}[c]{@{}c@{}}SBP\\ (mmHg)\end{tabular}}} & \multicolumn{2}{c|}{\textbf{\begin{tabular}[c]{@{}c@{}}DBP\\ (mmHg)\end{tabular}}} & \multicolumn{2}{c}{\textbf{\begin{tabular}[c]{@{}c@{}}SBP\\ (mmHg)\end{tabular}}} & \multicolumn{2}{c|}{\textbf{\begin{tabular}[c]{@{}c@{}}DBP\\ (mmHg)\end{tabular}}} & \multicolumn{2}{c|}{\textbf{\begin{tabular}[c]{@{}c@{}}SBP\\ (mmHg)\end{tabular}}} & \multicolumn{2}{c}{\textbf{\begin{tabular}[c]{@{}c@{}}DBP\\ (mmHg)\end{tabular}}} \\ \cline{2-13} 
\multicolumn{1}{c|}{} & \textbf{MAE} & \multicolumn{1}{c|}{\textbf{RMSE}} & \textbf{MAE} & \multicolumn{1}{c|}{\textbf{RMSE}} & \textbf{MAE} & \multicolumn{1}{c|}{\textbf{RMSE}} & \textbf{MAE} & \multicolumn{1}{c|}{\textbf{RMSE}} & \textbf{MAE} & \multicolumn{1}{c|}{\textbf{RMSE}} & \textbf{MAE} & \textbf{RMSE} \\ \hline
\multicolumn{1}{c|}{P-AdaBoost \cite{kachuee2017cuffless}} & 7.4185 & \multicolumn{1}{c|}{15.4952} & 3.5622 & \multicolumn{1}{c|}{7.7522} & $-$ & \multicolumn{1}{c|}{$-$} & $-$ & \multicolumn{1}{c|}{$-$} & $-$ & \multicolumn{1}{c|}{$-$} & $-$ & $-$ \\
\multicolumn{1}{c|}{W-AdaBoost \cite{kachuee2017cuffless}} & 9.4432 & \multicolumn{1}{c|}{17.7051} & 4.1831 & \multicolumn{1}{c|}{7.8119} & $-$ & \multicolumn{1}{c|}{$-$} & $-$ & \multicolumn{1}{c|}{$-$} & $-$ & \multicolumn{1}{c|}{$-$} & $-$ & $-$ \\
\multicolumn{1}{c|}{DeepRNN-4L \cite{su2017predicting}} & $-$ & \multicolumn{1}{c|}{$-$} & $-$ & \multicolumn{1}{c|}{$-$} & 1.8051 & \multicolumn{1}{c|}{2.8065} & 1.1026 & \multicolumn{1}{c|}{1.6832} & 1.5738 & \multicolumn{1}{c|}{2.1443} & 0.9518 & 1.3448 \\
\multicolumn{1}{c|}{Proposed Model} & $-$ & \multicolumn{1}{c|}{$-$} & $-$ & \multicolumn{1}{c|}{$-$} & \textbf{1.0993} & \multicolumn{1}{c|}{\textbf{1.5602}} & \textbf{0.5823} & \multicolumn{1}{c|}{\textbf{0.8488}} & \textbf{0.9283} & \multicolumn{1}{c|}{\textbf{1.2630}} & \textbf{0.5171} & \textbf{0.7280} \\ \hline
\end{tabular}
\centering
\end{table*}
\begin{table*}[t]
\centering
\begin{tabular}{ccccccccccccc}
\multicolumn{13}{c}{\textbf{\begin{tabular}[c]{@{}c@{}}Table III\\ Performance Evaluation Based on the AAMI Standard\\ (ME $<$ 5 mmHg, SDE \textless 8 mmHg)\end{tabular}}} \\ \hline\hline
\multicolumn{1}{c|}{\multirow{3}{*}{\textbf{Model}}} & \multicolumn{4}{c|}{\textbf{\begin{tabular}[c]{@{}c@{}}Sequence\\ Length\\ M=1\end{tabular}}} & \multicolumn{4}{c|}{\textbf{\begin{tabular}[c]{@{}c@{}}Sequence\\ Length\\ M=10\end{tabular}}} & \multicolumn{4}{c}{\textbf{\begin{tabular}[c]{@{}c@{}}Sequence\\ Length\\ M=32\end{tabular}}} \\ \cline{2-13} 
\multicolumn{1}{c|}{} & \multicolumn{2}{c|}{\textbf{\begin{tabular}[c]{@{}c@{}}SBP\\ (mmHg)\end{tabular}}} & \multicolumn{2}{c|}{\textbf{\begin{tabular}[c]{@{}c@{}}DBP\\ (mmHg)\end{tabular}}} & \multicolumn{2}{c}{\textbf{\begin{tabular}[c]{@{}c@{}}SBP\\ (mmHg)\end{tabular}}} & \multicolumn{2}{c|}{\textbf{\begin{tabular}[c]{@{}c@{}}DBP\\ (mmHg)\end{tabular}}} & \multicolumn{2}{c|}{\textbf{\begin{tabular}[c]{@{}c@{}}SBP\\ (mmHg)\end{tabular}}} & \multicolumn{2}{c}{\textbf{\begin{tabular}[c]{@{}c@{}}DBP\\ (mmHg)\end{tabular}}} \\ \cline{2-13} 
\multicolumn{1}{c|}{} & \textbf{ME} & \multicolumn{1}{c|}{\textbf{SDE}} & \textbf{ME} & \multicolumn{1}{c|}{\textbf{SDE}} & \textbf{ME} & \multicolumn{1}{c|}{\textbf{SDE}} & \textbf{ME} & \multicolumn{1}{c|}{\textbf{SDE}} & \textbf{ME} & \multicolumn{1}{c|}{\textbf{SDE}} & \textbf{ME} & \textbf{SDE} \\ \hline
\multicolumn{1}{c|}{P-AdaBoost \cite{kachuee2017cuffless}} & 0.0097 & \multicolumn{1}{c|}{15.4952} & 0.0740 & \multicolumn{1}{c|}{7.7522} & $-$ & \multicolumn{1}{c|}{$-$} & $-$ & \multicolumn{1}{c|}{$-$} & $-$ & \multicolumn{1}{c|}{$-$} & $-$ & $-$ \\
\multicolumn{1}{c|}{W-AdaBoost \cite{kachuee2017cuffless}} & 0.1398 & \multicolumn{1}{c|}{17.7051} & 0.0253 & \multicolumn{1}{c|}{7.8119} & $-$ & \multicolumn{1}{c|}{$-$} & $-$ & \multicolumn{1}{c|}{$-$} & $-$ & \multicolumn{1}{c|}{$-$} & $-$ & $-$ \\
\multicolumn{1}{c|}{DeepRNN-4L \cite{su2017predicting}} & $-$ & \multicolumn{1}{c|}{$-$} & $-$ & \multicolumn{1}{c|}{$-$} & \textbf{0.0175} & \multicolumn{1}{c|}{2.8065} & 0.0231 & \multicolumn{1}{c|}{1.6832} & 0.0778 & \multicolumn{1}{c|}{2.1443} & 0.0046 & 1.3448 \\
\multicolumn{1}{c|}{Proposed Model} & $-$ & \multicolumn{1}{c|}{$-$} & $-$ & \multicolumn{1}{c|}{$-$} & 0.0249 & \multicolumn{1}{c|}{\textbf{1.5602}} & \textbf{0.0103} & \multicolumn{1}{c|}{\textbf{0.8488}} & \textbf{0.0159} & \multicolumn{1}{c|}{\textbf{1.2630}} & \textbf{0.0018} & \textbf{0.7280} \\ \hline
\multicolumn{13}{l}{}
\end{tabular}
\centering
\end{table*}
\subsubsection{Network Implementation}
After preparing the input and output for the model the data for each patient has been split to 70\% for training, 10\% for validation and 20\% for test. The data is prepared for short ($M$=10, extracted from 16s window or 2000 samples at 125 Hz) and long ($M$=32, extracted from 40s window or 5000 samples at 125 Hz) sequences and training is done for both sequences.
\\The ANN-LSTM network mainly consists of two stacked LSTM layers with time-distributed ANNs connected to them. The ANNs consist of one hidden layer with 128 neurons with ReLU activation function. The number of hidden states of the LSTM network is selected as 128. Batch size of 128 is used. For loss function, mean square error (MSE) is used and for gradient optimization, Adam optimizer \cite{kingma2014adam} is used with an initial learning rate of 0.001. For regularization, the $l_2$ norm of the gradient is constrained to 3 and 5 for $M$ = 10 and $M$ = 32, respectively. The first LSTM layer is made bidirectional \cite{schuster1997bidirectional} to model the time variations in both directions. The primary objective of these fully connected networks is to extract the necessary features from the feature vector sequence ${\begin{bmatrix}\textbf{x}_1^T~\textbf{x}_2^T~\cdot\cdot\cdot~\textbf{x}_i^T~\cdot\cdot\cdot~\textbf{x}_M^T\end{bmatrix}}^T$, where each $\textbf{x}_i$ is a $1\times 513$ feature vector. The stacked LSTM layers learn the variation of the extracted features on different temporal scales to improve the BP estimation. The 2nd LSTM layer then provides the output as a sequence of vectors ${\begin{bmatrix}\textbf{y}_1^T~\textbf{y}_2^T~\cdot\cdot\cdot~\textbf{y}_i^T~\cdot\cdot\cdot~\textbf{y}_M^T\end{bmatrix}}^T$ where each $\textbf{y}_i$ is a $1 \times 2$ vector containing the estimated SBP and DBP. For continuous prediction, the result of $\textbf{y}_M$ is of interest.

\section{Results and Discussion}
The proposed model has been evaluated using mean absolute error (MAE) and root mean square error (RMSE) as metrics. MAE and RMSE are defined as
\begin{equation}
MAE=\frac{1}{N} \sum_{i=1}^{N} \mid z_{i_{M}} - y_{i_{M}}\mid
\end{equation}
and
\begin{equation}
RMSE=\frac{1}{N} \sqrt{ \sum_{i=1}^{N} \mid z_{i_{M}} - y_{i_{M}}\mid ^{2}}
\end{equation}
respectively, where $y_{M}$ and $z_{M}$ are the ground truth and estimated BP (SBP or DBP) for $M$th element of the time sequence, respectively. The performance of the proposed model is evaluated on AAMI and BHS standards. Bland-Altman analysis, Pearson's correlation coefficient analysis and box plot analysis is also performed on the proposed model. The results are reported in the following.

\subsection{Comparison with Other Works}
Table II shows the comparative analysis of different neural network architectures. From Table II, it can be seen that the adaptive boosting based models have acceptable error for DBP estimation, but high error for SBP estimation. The LSTM based models are able to model the variations in the features and thus the accuracy improves. The proposed ANN-LSTM model performs better than the rest of the models in terms of MAE and RMSE. As the sequence length $M$ is increased from 10 to 32 for both models, the prediction capability of the LSTM models increases. From Table II, it is also evident that all the models, including the proposed model, have shown lesser accuracy for SBP than DBP estimation. This lower accuracy of SBP estimation is directly related to the high variance of SBP data. Many of the subjects in the database have isolated systolic hypertension (ISH) \cite{staessen1990isolated}. ISH is a type of hypertension where the DBP of the patient is at normal range ($<$90 mmHg) but the SBP is abnormally high ($>$140 mmHg). Elderly patients are more prone to have ISH. The exact opposite case is isolated diastolic hypertension (IDH) \cite{pickering2003isolated}, where the SBP is normal ($<$140 mmHg) but the DBP is abnormally high ($>$90 mmHg). The used database does not contain any case of IDH. The inclusion of ISH patients increase the SD of SBP and it becomes harder for the models to learn the estimation of SBP. Thus, the error of the SBP estimation increases.

\begin{table*}[t]
\centering
\begin{tabular}{ccccccccccc}
\multicolumn{11}{c}{\textbf{\begin{tabular}[c]{@{}c@{}}Table IV\\ SBP Performance Evaluation Based on the BHS Standard\\ Minimum Cumulative Frequency (\%) of Error\\ Grade A - 60\% ($<$ 5 mmHg), 85\% ($<$ 10 mmHg), 95\% ($<$ 15 mmHg)\\ Grade B - 50\% ($<$ 5 mmHg), 75\% ($<$ 10 mmHg), 90\% ($<$ 15 mmHg)\\ Grade C - 40\% ($<$ 5 mmHg), 65\% ($<$ 10 mmHg), 85\% ($<$ 15 mmHg)\end{tabular}}} \\ \hline\hline
\multicolumn{1}{c|}{\multirow{2}{*}{\textbf{Model}}} & \multicolumn{1}{c|}{\multirow{2}{*}{\textbf{BP Type}}} & \multicolumn{3}{c|}{\textbf{\begin{tabular}[c]{@{}c@{}}Sequence\\ Length\\ M=1\end{tabular}}} & \multicolumn{3}{c|}{\textbf{\begin{tabular}[c]{@{}c@{}}Sequence\\ Length\\ M=10\end{tabular}}} & \multicolumn{3}{c}{\textbf{\begin{tabular}[c]{@{}c@{}}Sequence\\ Length\\ M=32\end{tabular}}} \\ \cline{3-11} 
\multicolumn{1}{c|}{} & \multicolumn{1}{c|}{} & \textbf{\begin{tabular}[c]{@{}c@{}}\textless 5\\ mmHg\\ (\%)\end{tabular}} & \textbf{\begin{tabular}[c]{@{}c@{}}\textless 10\\ mmHg\\ (\%)\end{tabular}} & \multicolumn{1}{c|}{\textbf{\begin{tabular}[c]{@{}c@{}}\textless 15\\ mmHg\\ (\%)\end{tabular}}} & \textbf{\begin{tabular}[c]{@{}c@{}}\textless 5\\ mmHg\\ (\%)\end{tabular}} & \textbf{\begin{tabular}[c]{@{}c@{}}\textless 10\\ mmHg\\ (\%)\end{tabular}} & \multicolumn{1}{c|}{\textbf{\begin{tabular}[c]{@{}c@{}}\textless 15\\ mmHg\\ (\%)\end{tabular}}} & \textbf{\begin{tabular}[c]{@{}c@{}}\textless 5\\ mmHg\\ (\%)\end{tabular}} & \textbf{\begin{tabular}[c]{@{}c@{}}\textless 10\\ mmHg\\ (\%)\end{tabular}} & \textbf{\begin{tabular}[c]{@{}c@{}}\textless 15\\ mmHg\\ (\%)\end{tabular}} \\ \hline
\multicolumn{1}{c|}{\multirow{2}{*}{P-AdaBoost \cite{kachuee2017cuffless}}} & \multicolumn{1}{c|}{SBP} & 66.76 & 82.29 & \multicolumn{1}{c|}{87.93} & $-$ & $-$ & \multicolumn{1}{c|}{$-$} & $-$ & $-$ & $-$ \\
\multicolumn{1}{c|}{} & \multicolumn{1}{c|}{DBP} & 83.12 & 91.06 & \multicolumn{1}{c|}{94.22} & $-$ & $-$ & \multicolumn{1}{c|}{$-$} & $-$ & $-$ & $-$ \\
\multicolumn{1}{c|}{\multirow{2}{*}{W-AdaBoost \cite{kachuee2017cuffless}}} & \multicolumn{1}{c|}{SBP} & 56.80 & 75.46 & \multicolumn{1}{c|}{83.11} & $-$ & $-$ & \multicolumn{1}{c|}{$-$} & $-$ & $-$ & $-$ \\
\multicolumn{1}{c|}{} & \multicolumn{1}{c|}{DBP} & 77.45 & 88.81 & \multicolumn{1}{c|}{93.35} & $-$ & $-$ & \multicolumn{1}{c|}{$-$} & $-$ & $-$ & $-$ \\
\multicolumn{1}{c|}{\multirow{2}{*}{DeepRNN-4L \cite{su2017predicting}}} & \multicolumn{1}{c|}{SBP} & $-$ & $-$ & \multicolumn{1}{c|}{$-$} & 94.82 & 99.39 & \multicolumn{1}{c|}{99.82} & 99.23 & 99.96 & 99.99 \\
\multicolumn{1}{c|}{} & \multicolumn{1}{c|}{DBP} & $-$ & $-$ & \multicolumn{1}{c|}{$-$} & 98.51 & 99.83 & \multicolumn{1}{c|}{99.93} & 99.80 & 99.98 & 99.99 \\
\multicolumn{1}{c|}{\multirow{2}{*}{Proposed Model}} & \multicolumn{1}{c|}{SBP} & $-$ & $-$ & \multicolumn{1}{c|}{$-$} & \textbf{98.98} & \textbf{99.92} & \multicolumn{1}{c|}{\textbf{99.98}} & \textbf{99.60} & \textbf{99.98} & \textbf{99.99} \\
\multicolumn{1}{c|}{} & \multicolumn{1}{c|}{DBP} & $-$ & $-$ & \multicolumn{1}{c|}{$-$} & \textbf{99.86} & \textbf{99.98} & \multicolumn{1}{c|}{\textbf{100.00}} & \textbf{99.96} & \textbf{99.99} & \textbf{100.00} \\ \hline
\end{tabular}
\centering
\end{table*}
\subsection{Performance Evaluation on AAMI Standards}
The Association for the Advancement of the Medical Instrumentation (AAMI) standard requires BP measurement devices to have mean error (ME) and standard deviation of error (SDE) values lower than 5 and 8 mmHg, respectively \cite{association1987american}. Table III presents an evaluation of different models according to the AAMI standard. It can be seen from Table III that only the LSTM-based models satisfy the AAMI standard for SBP estimation with the proposed model having the negligible ME and minimum SDE.
\begin{figure}[t]
\centering
\includegraphics[width=\linewidth]{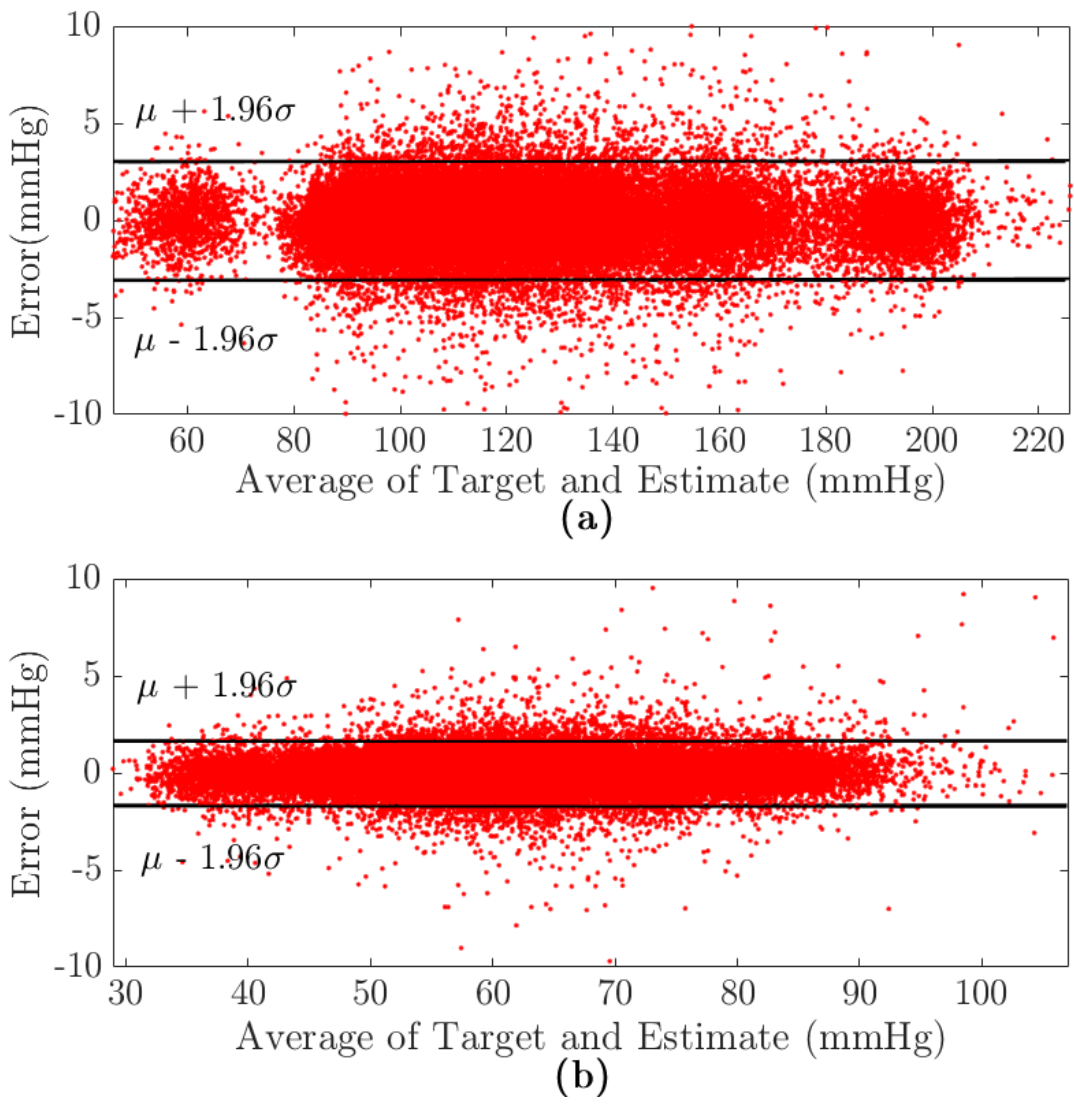}
\centering
\caption{Bland-Altman plot of (a) SBP and (b) DBP. The limits of agreement (LOA) for SBP and DBP are [$-$3.083,3.033] and [$-$1.674,1.653], respectively.}
\label{fig08_blandaltman}
\end{figure}
\begin{figure}[t]
\centering
\vspace*{0.15cm}
\includegraphics[width=\linewidth]{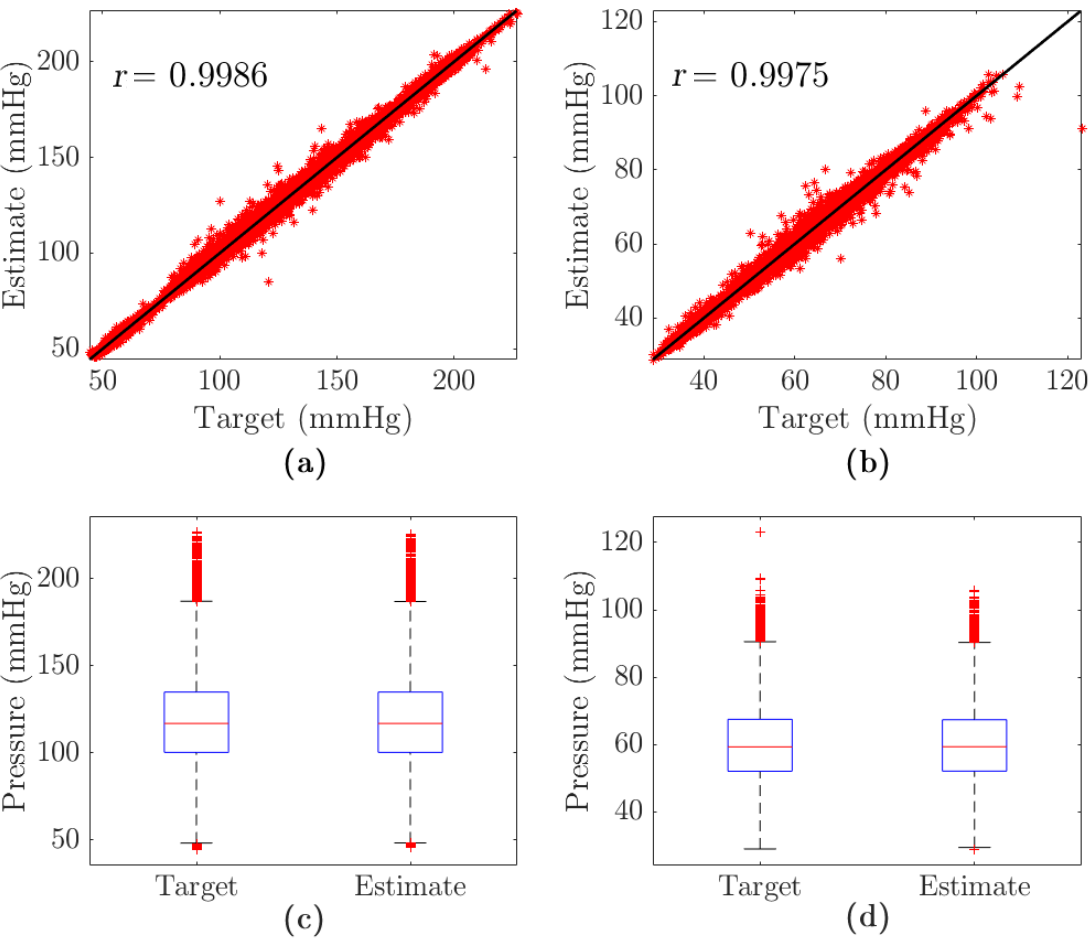}
\centering
\caption{The regression plot for (a) SBP and (b) DBP. The Pearson's correlation coefficient are $r$ = 0.9986 and $r$ = 0.9975 for SBP and DBP, respectively. Here, (c) and (d) shows the box plot for SBP and DBP. The high amount of outliers is due to the high SBP and DBP of some of the subjects. The similarity between the target and estimate box plots indicate that the proposed method is able to model the very high SBPs and DBPs.}
\label{fig09_regressionboxplot}
\end{figure}
\subsection{Performance Evaluation on BHS Standards}
Table IV presents an evaluation of different algorithms based on the British Hypertension Society (BHS) standard. BHS grades BP measurement devices based on their cumulative frequency percentage of errors less than three different thresholds, i.e., 5, 10, and 15 mmHg \cite{o1990british}. From Table IV, it is evident that the proposed model provides the best performance, with more than 98\% of the test samples having estimation errors less than 5 mmHg for both SBP and DBP.
\begin{figure*}[t]
\centering
\includegraphics[width=\linewidth]{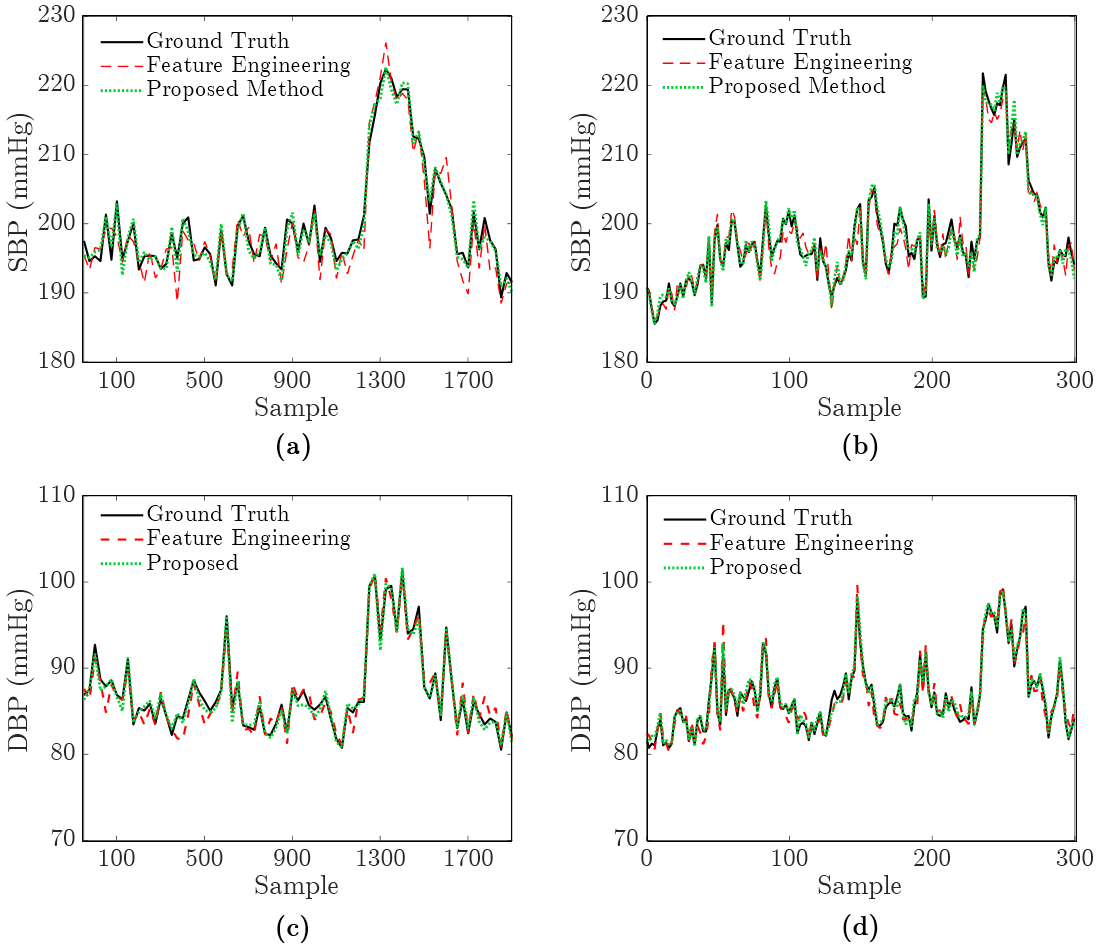}
\centering
\caption{Continuous BP measurement. Here, (a) and (c) are SBP and DBP measurement for $M$=10 and (b) and (d) are SBP and DBP measurement for $M$=32.}
\label{fig10_bptracking}
\end{figure*}

\subsection{Statistical Analysis}
Bland-Altman test \cite{altman1983measurement}, Pearson's correlation test and box plot analysis have also been performed on the data for short sequences ($M$ = 10). Fig. \ref{fig08_blandaltman} shows the Bland-Altman plot of SBP and DBP estimation. For the proposed model, the limits of agreement [$\mu - 1.96\sigma$, $\mu + 1.96\sigma$] for SBP and DBP have been found to be $[-3.083, 3.033]$ and $[-1.674, 1.653]$ respectively. It means that 95\% of the estimated SBPs have error less than 3.1 mmHg and 95\% of the measured DBPs have error less than 1.7 mmHg, which indicates that the model is a good estimator.

Fig. \ref{fig09_regressionboxplot}(a) and \ref{fig09_regressionboxplot}(b) show the regression plot for SBP and DBP estimation. The Pearson correlation coefficients of SBPs and DBPs are $r$ = 0.9986 and $r$ = 0.9975, respectively. Both of the coefficients are very close to 1.0, which indicates high linearity between the target and estimated BP.

Fig.\ref{fig09_regressionboxplot}(c)
and \ref{fig09_regressionboxplot}(d) show the box plot of SBP and DBP. In Fig. \ref{fig09_regressionboxplot}(c) and \ref{fig09_regressionboxplot}, the very high and very low SBP and DBP values are shown as red signs. The target and estimate box plots for SBP and DBP do not show any distinguishable difference. Thus, it is evident that the proposed model is able to estimate very high and very low SBPs and DBPs with high accuracy.

Fig. \ref{fig10_bptracking} shows the continuous SBP and DBP tracking of a patient using short and long sequences using the proposed method and feature engineering-based method \cite{su2017predicting}. From this figure, it can be seen that the proposed model is able to track continuous SBP and DBP with better accuracy than feature engineering-based models.

\section{Conclusion}
In this paper, we have proposed a novel waveform–based ANN-LSTM model for continuous blood pressure estimation model using ECG and PPG waveforms. The model is able to extract the necessary features without requiring any feature engineering and is also able to learn the variations of the features with time. For the same data, the proposed model performs better than the conventional feature engineering based methods. The proposed model satisfies the AAMI standard of BP estimation. And according to the BHS standard, the BP estimation quality for both SBP and DBP have achieved grade A. The Bland-Altman analysis shows LOA to be within $\pm$5 mmHg for both SBP and DBP. The Pearson's correlation test has also shown correlation coefficients close to 1.0, which indicates high linearity between the ground truth and estimated BP. The box plot analysis shows that the model is able to  account for very high and very low SBP and DBP values. In future works, we would like to investigate the model's robustness with larger number and wider variety of subjects.

\bibliographystyle{IEEEtran}
\bibliography{references}

\end{document}